\documentclass[11pt,preprint]{aastex}
\usepackage{epsfig}
\usepackage{natbib}
\usepackage{graphicx}
\usepackage{slashbox}
\usepackage{multirow}
\usepackage{lscape}
\usepackage{mathrsfs,amssymb}
\usepackage{amsmath}

\newcommand       \cm           {\,{\rm cm}}

\newcommand       \erg          {\,{\rm erg}}

\newcommand       \K            {\,{\rm K}}

\newcommand       \s            {\,{\rm s}}

\newcommand       \simlt        {\lesssim}

\newcommand       \gtsim        {\gtrsim}

\newcommand       \mum          {\,{\rm \mu m}}

\newcommand       \Teff         {T_{\rm eff}}

\newcommand       \simali       {\sim\,}

\newcommand       \Iratio         {{\rm I}_{3.4}/{\rm I}_{3.3}}
\newcommand       \Aratio        {{\rm A}_{3.4}/{\rm A}_{3.3}}
\newcommand       \Aali           {{\rm A}_{3.4}}
\newcommand       \Aaro          {{\rm A}_{3.3}}
\countdef\decade=200
\advance\decade by \year
\countdef\hours=201
\advance\hours by \time
\divide\hours by 60
\countdef\mins=202
\advance\mins by \hours
\multiply\mins by 60
\multiply\hours by 100
\countdef\miltime=203
\advance\miltime by \hours
\advance\miltime by \time
\advance\miltime by -\mins

\shorttitle{On the Carriers of the UIE Features}
\title{
\vspace*{-2.0em}
{\normalsize\rm Accepted for publication in
               {\it The Astrophysical Journal}}\\
\vspace*{1.0em}
The Carriers of the Interstellar Unidentified Infrared Emission Features:
Constraints from the Interstellar C--H Stretching Features at 3.2--3.5$\mum$
}
\author{X.J.~Yang\altaffilmark{1,2},
            R.~Glaser\altaffilmark{3},
            Aigen Li\altaffilmark{2},
            and J.X.~Zhong\altaffilmark{1}}
\altaffiltext{1}{Department of Physics,
                      Xiangtan University,
                      411105 Xiangtan, Hunan Province, China;
                      {\sf xjyang@xtu.edu.cn, jxzhong@xtu.edu.cn}}
\altaffiltext{2} {Department of Physics and Astronomy,
                  University of Missouri,
                  Columbia, MO 65211, USA;
                  {\sf lia@missouri.edu}}
\altaffiltext{3} {Department of Chemistry,
                  University of Missouri,
                  Columbia, MO 65211, USA;
                  {\sf glaserr@missouri.edu}}

\begin{document}

\begin{abstract}
The unidentified infrared emission (UIE) features
at 3.3, 6.2, 7.7, 8.6, and 11.3$\mum$,
commonly attributed to
polycyclic aromatic hydrocarbon (PAH) molecules,
have been recently ascribed to
mixed aromatic/aliphatic organic nanoparticles.
More recently, an upper limit of $<$\,9\% was placed
on the aliphatic fraction
(i.e., the fraction of carbon atoms in aliphatic form)
of the UIE carriers
based on the observed intensities of
the 3.4$\mum$ and 3.3$\mum$ emission features
by attributing them to aliphatic and aromatic
C--H stretching modes, respectively,
and assuming $\Aali/\Aaro \approx 0.68$
derived from a small set of aliphatic and aromatic compounds,
where $\Aali$ and $\Aaro$ are respectively the band strengths
of the 3.4$\mum$ aliphatic and 3.3$\mum$ aromatic C--H bonds.
%
%
%
To improve the estimate of
the aliphatic fraction of the UIE carriers,
here we analyze 35 UIE sources which exhibit
both the 3.3$\mum$ and 3.4$\mum$ C--H features
and determine $\Iratio$,
the ratio of the power emitted from
the 3.4$\mum$ feature
to that from the 3.3$\mum$ feature.
We derive the median ratio to be
$\langle\Iratio\rangle\approx 0.12$.
We employ density functional theory 
to
compute $\Aali/\Aaro$
for a range of methyl-substituted PAHs.
%
The resulting $\Aali/\Aaro$ ratio
well exceeds $\simali$1.4,
with an average ratio of $\Aali/\Aaro\approx 1.76$.
By attributing the 3.4$\mum$ feature
{\it exclusively} to aliphatic C--H stretch
(i.e., neglecting anharmonicity and superhydrogenation),
we derive the fraction of C atoms
in aliphatic form from $\Iratio\approx 0.12$
and $\Aali/\Aaro\approx 1.76$ to be $\sim$\,2\%.
We therefore conclude that the UIE emitters are
predominantly aromatic.
\end{abstract}
\keywords {dust, extinction --- ISM: lines and bands --- ISM: molecules}

\section{Introduction\label{sec:intro}}
Many astronomical objects show a distinctive set of
emission features in the infrared (IR),
collectively known as the  ``unidentified infrared
emission'' (UIE) bands. The strongest bands fall at
3.3, 6.2, 7.7, 8.6 and 11.3$\mum$.
They are ubiquitously seen in a wide variety of astrophysical
environments, including galaxies at redshifts $z$\,$>$\,2.
They account for $>$\,10--20\% of the total IR power of the Milky Way
and star-forming galaxies (Tielens 2008).

Since their first detection four decades ago
(Gillett et al.\ 1973),
the carriers of the UIE bands remain unidentified.
A popular hypothesis is that the UIE bands arise from
the vibrational modes of gas-phase, free-flying
polycyclic aromatic hydrocarbon (PAH) molecules
which are predominantly aromatic
(L\'eger \& Puget 1984, Allamandola et al.\ 1985, 1989, 1999).
Alternatively, amorphous solids with
a mixed aromatic/aliphatic structure,
such as hydrogenated amorphous carbon (HAC; Jones et al.\ 1990),
quenched carbonaceous composite (QCC; Sakata et al.\ 1987),
and coal (Papoular et al.\ 1993) have also been proposed.

The HAC, QCC, and coal hypotheses assume that the UIE bands
arise following photon absorption in
small thermally-isolated aromatic units
within or attached to these bulk materials
(Duley \& Williams 1981).
However, it does not appear possible to confine
the absorbed stellar photon energy within these aromatic
``islands'' for the time $\gtsim10^{-3}\s$ required for
the thermal emission process (see Li \& Draine 2002).
Bulk materials like HAC, QCC and coal
have a huge number of vibrational degrees of freedom and
therefore their heat capacities are so large that they will
attain an equilibrium temperature of $T$\,$\simali$20$\K$
in the diffuse interstellar medium (ISM).
With $T$\,$\simali$20$\K$,
they will not emit efficiently in the UIE bands
at $\lambda$\,$\simali$3--12$\mum$ (see Li 2004).

Recognizing the challenge of
bulk materials like HAC, QCC and coal
in being heated to emit the UIE bands,
Kwok \& Zhang (2011, 2013) recently proposed
the so-called MAON model: they argued that the UIE bands
arise from coal- or kerogen-like organic nanoparticles,
consisting of chain-like aliphatic hydrocarbon material
linking small units of aromatic rings,
where MAON stands for ``{\it mixed aromatic/aliphatic
organic nanoparticle}''.
The major improvement of the MAON model
over the earlier HAC, QCC and coal hypotheses is that
the MAON model hypothesizes
that the coal-like UIE carriers are {\it nanometer} in size
so that their heat capacities are smaller than or comparable
to the energy of the starlight photons that excite them.
Upon absorption of a single stellar photon,
they will be stochastically heated to high temperatures
to emit the UIE bands (see Draine \& Li 2001).

In brief, the current views about the UIE carriers
generally agree that (1) the UIE features arise from
the {\it aromatic} C--C and C--H vibrational modes,
and (2) the carriers must be {\it nanometer} in size or smaller
(e.g., large molecules). The dispute is mainly on the structure
of the UIE carriers:
are they predominantly {\it aromatic} (like PAHs),
or {\it largely aliphatic}
but mixed with small aromatic units (like MAONs)?

Are the UIE carriers aromatic or aliphatic?
A straightforward way to address this question
is to examine the {\it aliphatic fraction}
of the UIE carriers
(i.e., the fraction of carbon atoms in aliphatic chains).
Aliphatic hydrocarbons have a vibrational
band at 3.4$\mum$ due to the C--H stretching mode
(Pendleton \& Allamandola 2002).
In some HII regions, reflection nebulae and planetary nebulae
the UIE band near 3$\mum$ exhibits a rich spectrum:
the dominant 3.3$\mum$ feature is usually accompanied
by a weaker feature at 3.4$\mum$
along with an underlying plateau
extending out to $\simali$3.6$\mum$.
In some objects, a series of weaker features
at 3.46, 3.51, and 3.56$\mum$ are also seen superimposed
on the plateau, showing a tendency to decrease in strength
with increasing wavelength
(see Geballe et al.\ 1985, Jourdain de Muizon et al.\ 1986,
Joblin et al.\ 1996).
While the assignment of the 3.3$\mum$ emission feature to
the aromatic C--H stretch is widely accepted,
the precise identification of the 3.4$\mum$ feature
(and the accompanying weak features at 3.46, 3.51, and 3.56$\mum$
and the broad plateau) remains somewhat controversial.
By assigning the 3.4$\mum$ emission {\it exclusively}
to aliphatic C--H, one can place an upper limit
on the aliphatic fraction of the emitters
of the UIE features.
This is indeed an {\it upper limit}
as the 3.4$\mum$ emission feature could also be
due to {\it anharmonicity} of the aromatic C--H
stretch (Barker et al.\ 1987) 
and ``{\it superhydrogenated}'' PAHs
whose edges contain excess H atoms
(Bernstein et al.\ 1996, Sandford et al.\ 2013).

Let ${\rm I}_{3.4}$ and ${\rm I}_{3.3}$ respectively
be the observed intensities of the 3.4$\mum$
and 3.3$\mum$ emission features.
Let ${\rm A}_{3.4}$ and ${\rm A}_{3.3}$ respectively
be the band strengths 
(on a per unit C--H bond basis)
of the aliphatic and aromatic C--H bonds.
Let $N_{\rm H,aliph}$ and $N_{\rm H,arom}$ respectively
be the numbers of aliphatic and aromatic C--H bonds
in the emitters of the 3.3$\mum$ UIE feature.
We obtain
$N_{\rm H,aliph}/N_{\rm H,arom}\approx\left({\rm I}_{3.4}/{\rm I}_{3.3}\right)
\times\left({\rm A}_{3.3}/{\rm A}_{3.4}\right)$.
We assume that one aliphatic C atom corresponds to
2.5 aliphatic C--H bonds (intermediate between methylene --CH$_2$
and methyl --CH$_3$) and one aromatic C atom corresponds to
0.75 aromatic C--H bond (intermediate between benzene C$_6$H$_6$
and coronene C$_{24}$H$_{12}$).
Therefore, in the UIE carriers the ratio of the number of C atoms
in aliphatic units to that in aromatic rings is
$N_{\rm C,aliph}/N_{\rm C,arom}\approx
\left(0.75/2.5\right)\,\times\,N_{\rm H,aliph}/N_{\rm H,arom}
= 0.3\times\,\left({\rm I}_{3.4}/{\rm I}_{3.3}\right)
\times\,\left({\rm A}_{3.3}/{\rm A}_{3.4}\right)$.

In \S\ref{sec:Iratio} we compile the UIE spectra of
a wide variety of objects and determine $\Iratio$,
the ratio of the power emitted from
the 3.4$\mum$ aliphatic C--H feature
to that from the 3.3$\mum$ aromatic C--H feature.
In \S\ref{sec:Aratio} we use
density functional theory (DFT) 
to calculate the band strengths 
of the 3.4$\mum$ feature ($\Aali$)
and the 3.3$\mum$ feature ($\Aaro$)
for a range of PAH molecules
with a methyl side chain.
We estimate in \S\ref{sec:astro} the aliphatic fraction
of the UIE carriers from $\Iratio$ and $\Aratio$.
We summarize our major results in \S\ref{sec:summary}.

\begin{figure}
\centerline
{
\includegraphics[width=12cm,angle=0]{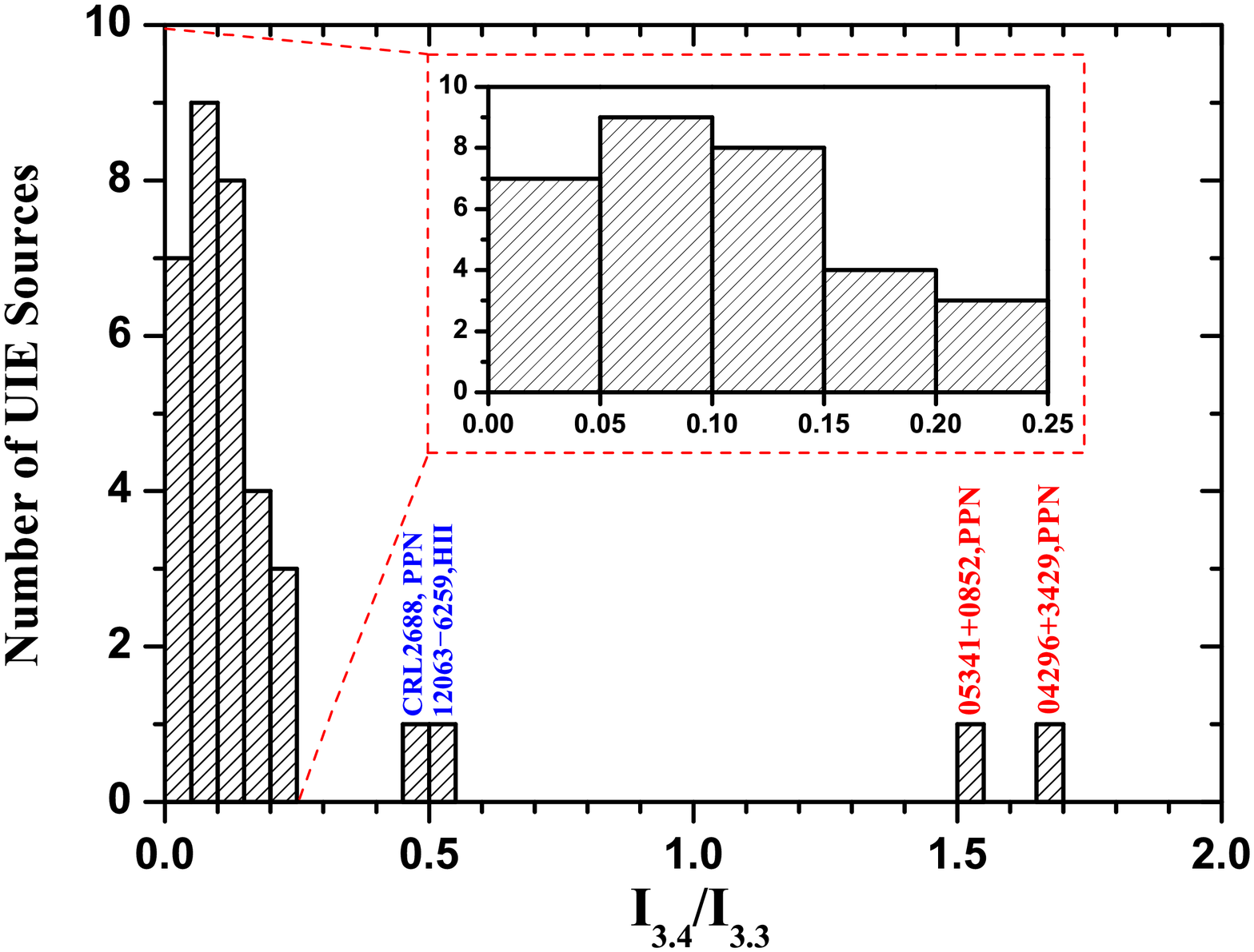}
}
\caption{\footnotesize
         \label{fig:Iratio}
         Histogram of the flux ratio ($\Iratio$) for 35 UIE sources.
         The median flux ratio is $\langle \Iratio\rangle\approx 0.12$.
         The insert panel enlarges the flux ratio distribution for the
         31 sources with $\Iratio\simlt 0.25$.
         }
\end{figure}

\section{$\Iratio$: Flux Ratio of
         the 3.4$\mum$ Aliphatic C--H Feature
         to the 3.3$\mum$ Aromatic C--H Feature
         \label{sec:Iratio}
         }
We compile all the UIE spectra available in the literature
which exhibit both the 3.3$\mum$ aromatic C--H feature
and the 3.4$\mum$ aliphatic C--H feature.
We fit the observed spectra in terms of
Drude profiles combined with an underlying continuum.
The Drude profile,
expected for classical damped harmonic oscillators,
is characterized by
$\lambda_{\rm o}$ -- the peak wavelength,
$\gamma$ -- the band width,
and $F_{\rm int}$ -- the integrated flux
of the emission feature:
$F_\lambda = F_{\rm int}\,\times\,\left(2\gamma\right/\pi)/\left[\left(\lambda-\lambda_{\rm o}^2/\lambda\right)^2+\gamma^2\right]$,
where $F_{\rm int}=\int F_\lambda d\lambda$,
and $F_\lambda$ is the flux at wavelength $\lambda$
in unit of $\erg\s^{-1}\cm^{-2}\mum^{-1}$.
The ratio of the power emitted from the 3.4$\mum$ feature
to that from the 3.3$\mum$ feature is
$\Iratio = F_{\rm int}(3.4)/F_{\rm int}(3.3)$.

We note that in the literature the band ratios $\Iratio$
have been reported for some sources.
Nevertheless, we prefer to derive the band ratios
by ourselves because the previous analyses used
different methods to calculate the strengths of
the features and the definitions of the underlying
continuum of the features were somewhat arbitrary.
These differences are not negligible in comparison
with data reported in different papers.
We therefore decide to derive $\Iratio$
in a consistent way for all sources.
%

The above-described Drude fitting has been performed
for 35 UIE sources, spanning a wide range of environments:
reflection nebulae, HII regions, molecular clouds,
photodissociated regions, planetary nebulae and
protoplanetary nebulae. In Figure~\ref{fig:Iratio} we show
the flux ratio ($\Iratio$) distribution of these sources.
It can be seen that the majority (31/35) of these sources
has $\Iratio < 0.25$. The median of the $\Iratio$ ratio is
$\langle \Iratio\rangle\approx 0.12$.
Two protoplanetary nebulae
(IRAS\,04296+3429, IRAS\,05341+0852)
have an unusually high $\Iratio$ ratio
and will be discussed elsewhere.

\begin{figure}[ht]
 \vspace{-2mm}
  \begin{center}
  \epsfig{file=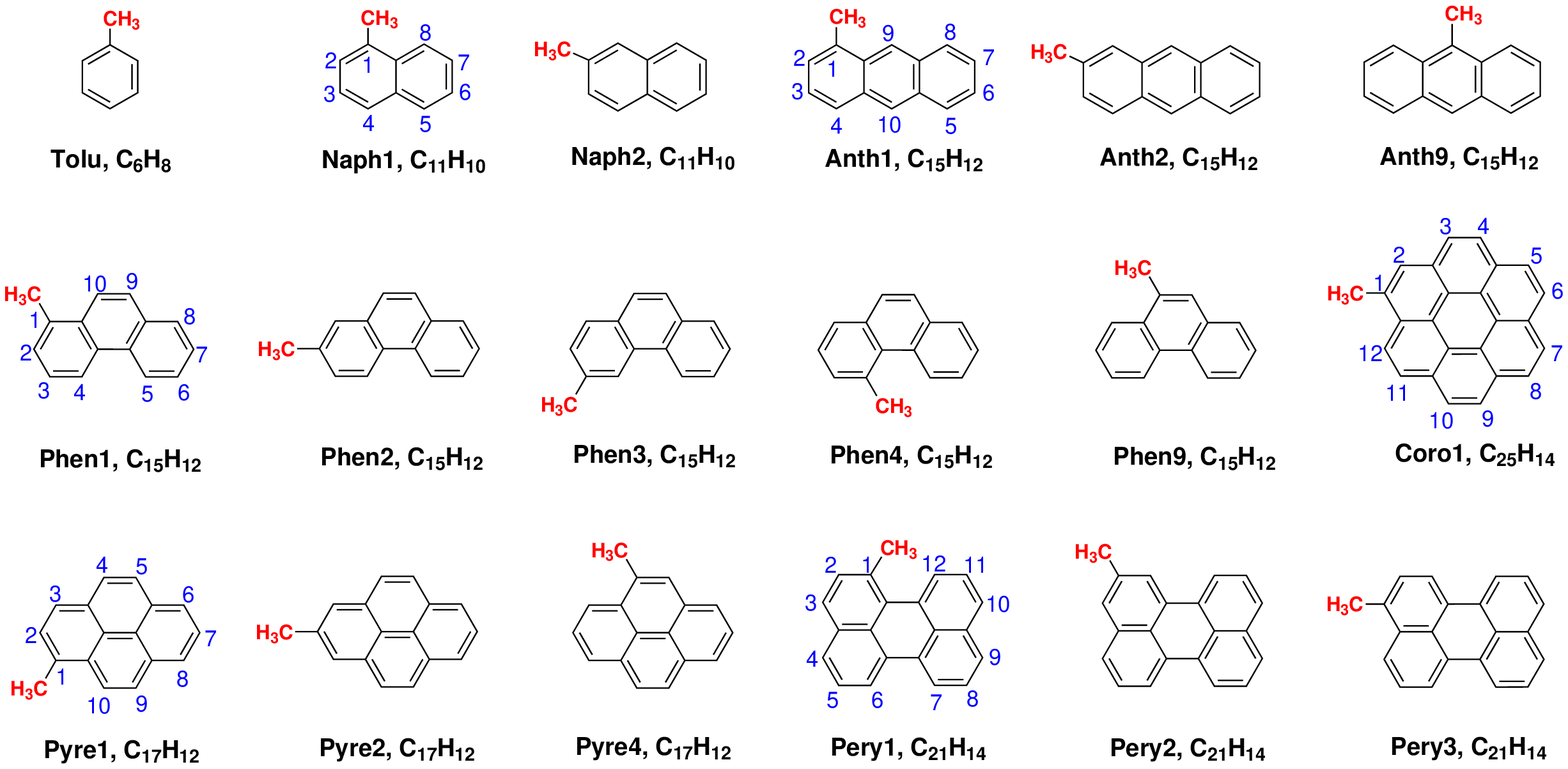,width=13.6cm}
  \end{center}
\vspace{-6mm}
\caption{\label{fig:MonoMethylPAHs} \footnotesize
         Structures of the mono-methyl (${\rm -CH_3}$) derivatives
         of seven aromatic molecules together with the standard 
         {\it International Union of Pure and Applied Chemistry}
         (IUPAC) numbering:
         benzene (C$_6$H$_6$),
         naphthalene (C$_{10}$H$_8$),
         anthracene (C$_{14}$H$_{10}$),
         phenanthrene (C$_{14}$H$_{10}$),
         pyrene (C$_{16}$H$_{10}$),
         perylene (C$_{20}$H$_{12}$), and
         coronene (C$_{24}$H$_{12}$).
         We use the first four letters of the parent molecules to
         refer to them and attach the position number of
         the location of the methyl group
         (e.g., Naph1 for 1-methylnaphthalene).
         The mono-methyl derivative of benzene
         is known as toluene (i.e., ``Tolu'', C$_7$H$_8$).
         Depending on where the methyl side-group
         is attached, a molecule will have several
         isomers (e.g., monomethyl-pyrene has three isomers
         in which the ${\rm -CH_3}$ side-group is respectively
         attached to the edge position 1, 2, and 4).
	 }
\vspace{-3mm}
\end{figure}

\begin{figure}
\vspace{-10mm}
\centerline
{
\includegraphics[width=13.6cm,angle=0]{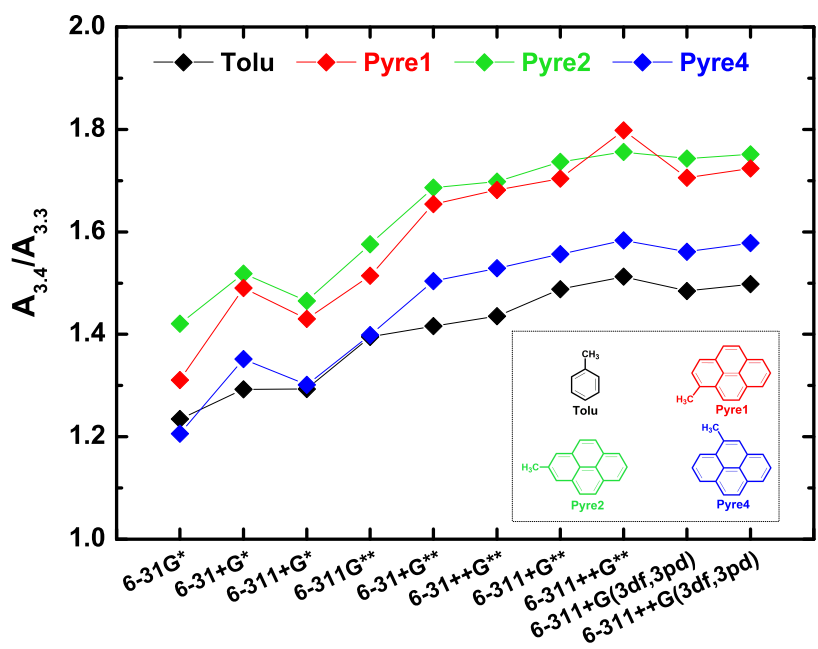}
}
\vspace{-4mm}
\caption{\footnotesize
         \label{BasisSet}
         Band-strength ratios $\Aratio$ computed with 
         different basis sets for toluene (i.e., methyl-benzene)
         and the three isomers of methyl-pyrene.
         From left to right, the computations become
         increasingly more computertime-intensive and the results are
         expected to be more accurate.
         The results computed with the B3LYP method and
         in conjunction with the basis sets
         {\rm 6-311+G$^{\ast\ast}$},
         {\rm 6-311++G$^{\ast\ast}$},
         {\rm 6-311+G(3df,3pd)}, and
         6-311++G(3df,3pd)
         have essentially reached the convergence limit.
         For a compromise between accuracy
         and computational demand, we adopt
         the method of {\rm B3LYP/6-311+G$^{\ast\ast}$}.
         }
\end{figure}

\begin{figure}
\centerline
{
\includegraphics[width=16cm,angle=0]{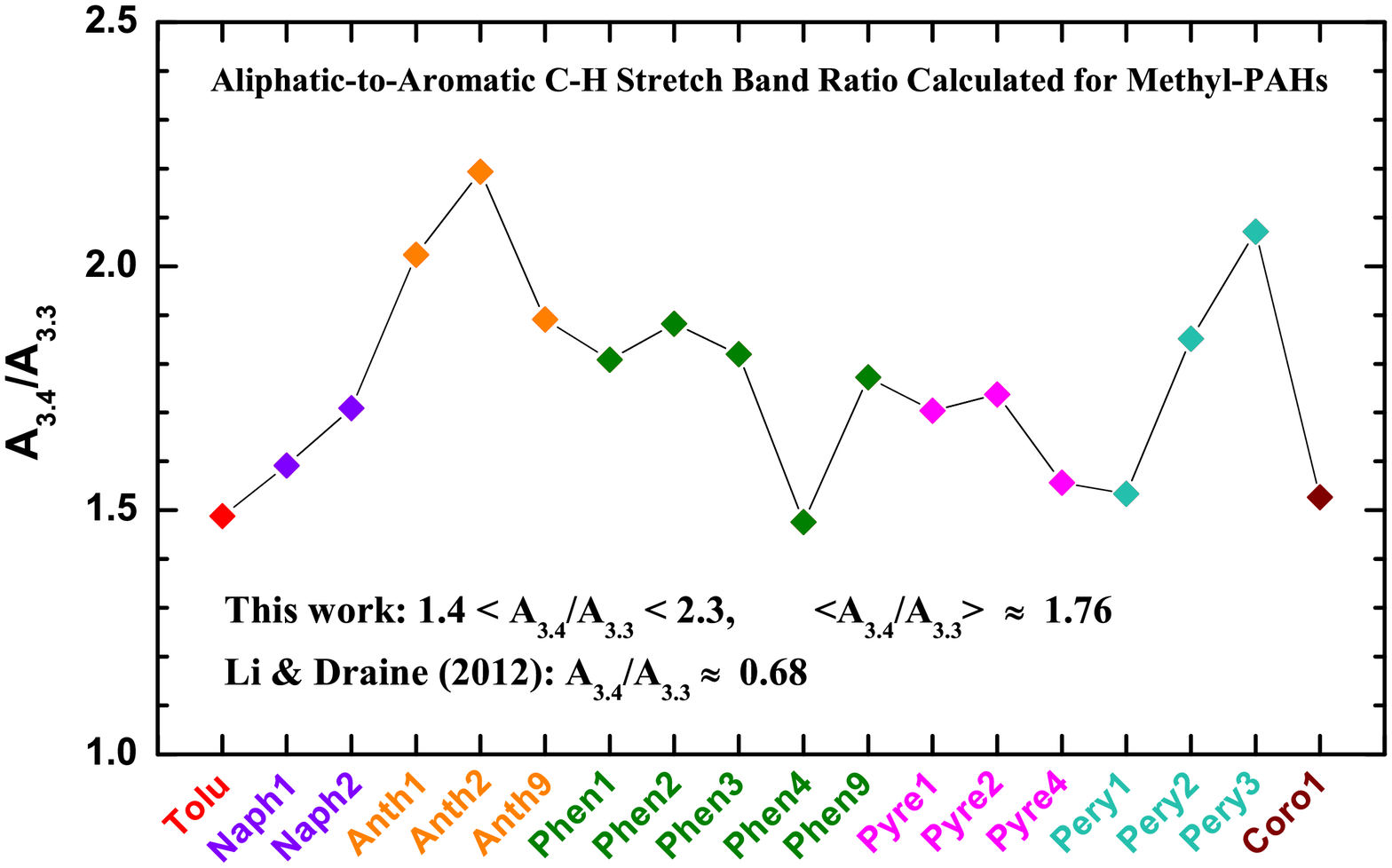}
}
\caption{\footnotesize
         \label{fig:A34A33}
         Band-strength ratios as determined
         with the B3LYP/{\rm 6-311+G$^{\ast\ast}$} method
         for the mono-methyl derivatives
         of seven aromatic molecules and all of their isomers
         (benzene, naphthalene, anthracene,
         phenanthrene, pyrene, perylene, and coronene;
         see Figure~\ref{fig:MonoMethylPAHs}).
         The band-strength ratios $\Aratio$
         are all above $\simali$1.4, with a mean ratio of
         $\langle \Aratio\rangle \approx 1.76$.
         }
\end{figure}

\section{$\Aratio$: Band-Strength Ratio of
         the 3.4$\mum$ Aliphatic C--H Feature
         to the 3.3$\mum$ Aromatic C--H Feature
         \label{sec:Aratio}
         }
The 3.4$\mum$ feature may result from aliphatic side chains
attached as functional groups to PAHs
(Duley \& Williams 1981, Pauzat et al.\ 1999, Wagner et al.\ 2000).
The C--H stretch of methyl (--CH$_3$),
methylene (--CH$_2$--), and ethyl (--CH$_2$CH$_3$)
side chains on PAHs falls near the weaker satellite features
associated with the 3.3$\mum$ band.

We use the Gaussian09 software (Frisch et al.\ 2009) to calculate
the band strengths of the 3.4$\mum$ feature ($\Aali$)
and the 3.3$\mum$ feature ($\Aaro$) for a range of
aromatic molecules with a methyl side chain
(Figure~\ref{fig:MonoMethylPAHs}).
We have considered seven molecules:
benzene (C$_6$H$_6$),
naphthalene (C$_{10}$H$_8$),
anthracene (C$_{14}$H$_{10}$),
phenanthrene (C$_{14}$H$_{10}$),
pyrene (C$_{16}$H$_{10}$),
perylene (C$_{20}$H$_{12}$), and
coronene (C$_{24}$H$_{12}$).
Depending on the position of the methyl attachment,
a specific aromatic molecule can have
several structural isomers
(e.g., the three isomers of methyl-pyrene
have the ${\rm -CH_3}$ group attached to
the edge position 1, 2, or 4, respectively;
see Figure~\ref{fig:MonoMethylPAHs}).

The methyl group is taken to represent the aliphatic
component of the UIE carriers.
We will focus on methyl-substituted PAHs
as PAHs with larger side chains
are not as stable against photolytic
dissociation as methyl-PAHs
and are therefore not expected to
present in a large abundance.\footnote{%
    If a large aliphatic chain (e.g., --CH$_2$--CH$_3$)
    is attached to an aromatic structure, the most likely
    photodissociation product is a benzyl radical
    PAH-$\dot{\rm C}$H$_2$
    (i.e., an --CH$_2$ group attached to a PAH molecule),
    which, when subjected to the reaction
    PAH-$\dot{\rm C}$H$_2$ + H $\rightarrow$ PAH--CH$_3$
    will rapidly lead to the product of a CH$_3$ side group
    at the periphery of an aromatic molecule
    (Joblin et al.\ 1996; Hwang et al.\ 2002).
    \label{ftnt:CH3}
    }


The relative abundance of isomers of
a given methyl-substituted PAH will depend
on the chemical pathways for their formation
(e.g., through the photodegradation of larger PAHs).
We also do not have any knowledge of the relative
abundance of the various PAHs. In the absence of
this information, the band-strength ratio $\Aratio$
will be derived by averaging over all the isomers
of the selected PAHs.

The computations can be performed at various theoretical levels
(Koch et al.\ 2001; Cramer et al.\ 2004; Sholl et al.\ 2009).
A theoretical level denotes a combination of a quantum-mechanical
method and a basis set describing the atomic orbitals.
In the present study, we primarily employ
the hybrid density functional theoretical method B3LYP.
Direct comparisons are made to experimental gas phase
IR spectra of benzene, toluene, Naph, Naph1, Naph2,
Anth, Anth1, Anth2, Anth9, Phen, Phen1, Phen2, Phen3,
Pyre, Pyre1 and Coro (see Figure~\ref{fig:MonoMethylPAHs}),
where the molecule is labeled by the first four letters
(e.g., Naph refers to naphthalene)
and the number refers to the position where the methyl side chain
attaches (e.g., Anth2 refers to 2-methylanthracene,
i.e., anthracene with a methyl side chain attached to position 2).
In this paper, we report the band strength ratios 
$\Aratio$ obtained with the B3LYP method since the $\Aratio$ 
ratios are all we really need for estimating the aliphatic
fractions of the UIR carriers.\footnote{%
  In order not to distract from the major
  astro-message, the extensive details on 
  computational aspects and results will be 
  reported in a separate paper
  (see X.J. Yang et al.\ 2013, in preparation), 
  including the frequency and intensity scaling,
  line assignment and interpretation,
  comparison with experimental data,
  and detailed (comparative) analysis
  of the line intensities (and their uncertainties)
  computed with different methods 
  and with different basis sets.
  }

The B3LYP calculations are performed with a variety of
basis sets starting with {\rm 6-31G$^{\ast}$}
and all the way up to 6-311++G(3df,3pd).
As shown in Figure~\ref{BasisSet}
the band-strength ratios $\Aratio$
computed with the basis sets
{\rm 6-311+G$^{\ast\ast}$},
{\rm 6-311++G$^{\ast\ast}$},
{\rm 6-311+G(3df,3pd)}, and
6-311++G(3df,3pd)
have essentially reached the convergence limit.
The {\rm B3LYP/6-311+G$^{\ast\ast}$} method presents
an excellent compromise between accuracy
and computational demand,
and we adopt this theoretical level to
compute all of the molecules
shown in Figure~\ref{fig:MonoMethylPAHs}.

To examine the accuracy of the B3LYP method,
we have also employed 
second-order M$\o$ller-Plesset perturbation theory (MP2)
to compute the band strength ratios $\Aratio$ 
for toluene, Naph1 and Naph2.
The MP2 method is thought to be more accurate 
in computing band intensities than B3LYP 
(see Cramer et al.\ 2004).
As shown in Table~\ref{tab:DFTvsMP2},
the $\Aratio$ ratios of toluene, Naph1 and Naph2
computed from B3LYP/6-311++G(3df,3pd)
closely agree with that from 
MP2/6-311++G(3df,3pd) within $\simlt$4\%. 
With the same basis set,
the MP2 method is computationally much more
expensive than B3LYP
and it is therefore not practical to use MP2 to compute
the $\Aratio$ ratios for all the considered molecules. 
Also shown in Table~\ref{tab:DFTvsMP2} are 
the $\Aratio$ ratios given by the experimental 
gas-phase spectra of toluene, Naph1 and Naph2
of the {\it National Institute of Standards and Technology} 
(NIST). 
The agreement between the $\Aratio$ ratios computed from
 B3LYP/{\rm 6-311+G$^{\ast\ast}$} 
and that from the {\it NIST} experimental spectra
is reasonably close:
the largest difference is for toluene
(which is only $\simali$13\%).

\begin{table*}
\caption[]{\footnotesize
           Comparison of the band strength ratios $\Aratio$ 
           of toluene, Naph1 and Naph2 computed
           from B3LYP with that from MP2 
           and with that from the {\it NIST} 
           experimental spectra 
           of these molecules
           (http://webbook.nist.gov).
           }
\label{tab:DFTvsMP2}
\centerline{
\begin{tabular}{lccc}
\noalign{\smallskip} \hline \hline \noalign{\smallskip}
  Method  &   Toluene     &  Naph1  &    Naph2 \\ 
\noalign{\smallskip} \hline \noalign{\smallskip}
B3LYP/6-311+G$^{\ast\ast}$  &   1.49 &  1.59  &  1.71 \\
MP2/6-311+G$^{\ast\ast}$    &   1.52 &  1.50  &  1.57 \\
\hline
B3LYP/6-311+G(3df,3pd)      &   1.48 &  1.59  &  1.70 \\
MP2/6-311+G(3df,3pd)        &   1.56 &  1.52  &  1.66 \\
\hline 
{\it NIST} Experimental     &   1.32 &  1.66  &  1.89 \\
\hline 
\hline
\noalign{\smallskip}
\noalign{\smallskip} \noalign{\smallskip}
\end{tabular}
}
\end{table*}


The absolute intensity of an aromatic C--H stretch
depends relatively little on the size and nature of
the PAH, this intensity is about
$14\pm2\,{\rm km}\,{\rm mol}^{-1}$
(i.e., 2.3$\pm{0.3}\times10^{-18}\cm$ per C--H bond).
In sharp contrast, the intensity of a methyl C--H stretch
depends greatly on the nature of the PAH
and also on the specific isomer.
For example, the methyl group of the five phenanthrene isomers
give rise to $\Aratio$ between 1.4 and 1.9.
A similar isomer dependency is observed for perylene
with $\Aratio$ between 1.5 and 2.2.
The average $\Aratio$ ratio determined for our set of molecules
is $\simali$1.76, and the individual ratios fall within the broad
range between 1.4 and 2.3
(Figure~\ref{fig:A34A33}).
It is important to fully realize this
high structure-dependency of the $\Aratio$ ratio and this finding
stresses the need to study the formation processes
for methyl-substituted PAHs.

\begin{figure}[ht]
 \vspace{-3mm}
  \begin{center}
\includegraphics[width=12.8cm,angle=0]{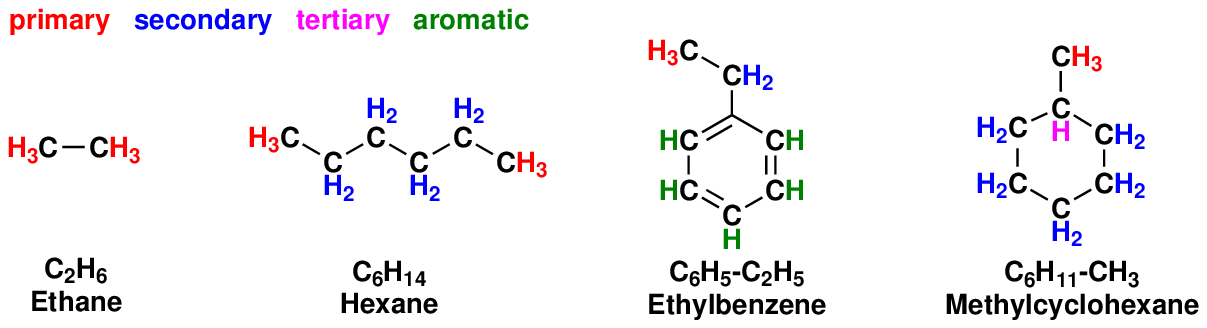}
  \end{center}
\vspace{-9mm}
\caption{\label{fig:LD12Mol} \footnotesize
         Structures of the molecules based on which
         Li \& Draine (2012) derived the 3.4$\mum$
         aliphatic C--H band strength ($\Aali$).
	 }
\vspace{-3mm}
\end{figure}

\section{Results and Discussion\label{sec:astro}}
%
%
Taking ${\rm I}_{3.4}/{\rm I}_{3.3}$\,=\,0.12
(\S\ref{sec:Iratio})
and $\Aratio$\,$\approx$\,1.76 (\S\ref{sec:Aratio}),
we estimate in the UIE carriers the ratio of
the number of C atoms in aliphatic units
to that in aromatic rings to be
$N_{\rm C,aliph}/N_{\rm C,arom}\approx
0.3\times\,\left({\rm I}_{3.4}/{\rm I}_{3.3}\right)
\times\,\left({\rm A}_{3.3}/{\rm A}_{3.4}\right)
\approx 0.02$.
Clearly, these results show in a compelling fashion that
the aliphatic component is only a very minor part of
the UIE emitters.

Li \& Draine (2012) examined the aliphatic fraction
of the UIE carriers in NGC\,7027 (a planetary nebula
with $\Iratio\approx 0.22$)
and the Orion bar (a photodissociated region
with $\Iratio\approx 0.19$).
They adopted a band-strength ratio of $\Aratio\approx 0.68$,
with $\Aaro = 4.0\times10^{-18}\cm$ per aromatic C--H bond
for small neutral PAHs (Draine \& Li 2007),
and ${\rm A}_{3.4} = 2.7\times10^{-18}\cm$ per aliphatic C--H bond,
averaged over ethane, hexane, ethyl-benzene,
and methyl-cyclo-hexane (d'Hendecourt \& Allamandola 1986;
see Figure~\ref{fig:LD12Mol}).
They placed an upper limit of $\simali$9\% on the aliphatic
fraction of the UIE carriers.\footnote{%
   Li \& Draine (2012) derived an upper limit of $\simali$15\%
   for the aliphatic fraction of the UIE emitters
   based on the C--H deformation band at 6.85$\mum$.
   }
We argue that the band-strength ratios $\Aratio$ computed
in \S\ref{sec:Aratio} are better suited for studying
the aliphatic fraction of the UIE carriers
as ethane, hexane, and methyl-cyclo-hexane
are pure aliphatic molecules.
It is known that the band strength of
the 3.4$\mum$ aliphatic C--H stretch
of pure aliphatics differs appreciably
from that of aromatics with aliphatic
side chains (see Wexler 1967).

We have studied mono-methyl derivatives of selected,
relatively small PAHs. In reality, one would assume
that the PAH molecules in space cover a much larger
range of sizes, from a few tens of C atoms up to several
thousands, with a mean size of $\simali$100 C atoms
(see Li \& Draine 2001).
They could include larger alkyl side chains
(ethyl, propyl, butyl, ...),
and several alkyl side chains might be present
in one PAH molecule.
Moreover, the alkyl side chains might be unsaturated
(i.e., --CH=CH$_2$, --CH=CH--, C=CH$_2$, C=C--H;
see Kwok \& Zhang 2013).
We have already argued that methyl side chains
are more likely than larger alkyl side chains
because of the photodissociation of the latter
(also see Footnote~\ref{ftnt:CH3}).

In preliminary studies, we have examined all possible
isomers of di-methyl-substituted pyrene and found that
the methyl groups are essentially independent of each other.
Noticeable effects on frequency and intensity only occur
when several alkyl groups are placed in direct proximity.
The study of larger PAHs obviously will be a target in future.
However, the finding of the strong isomer dependency
suggests that it will be more important to first learn
about the probabilities of the formations
of the various isomers.
The effects of unsaturated side chains on the IR spectra
remains to be examined.
While the positions of the C--H stretches of
simple alkenes and dienes coincide with
the methyl signals of methyl-substituted PAHs,
the aliphatic C--H stretches of styrene-type molecules 
(PAH--CH=CH$_2$) and of their derivatives fall in 
the wavelength range shortward of the aromatic 
C--H stretch at 3.3$\mum$ which are not seen in the ISM.
%

Kwok \& Zhang (2013) argued that the 3.4$\mum$ feature
may not be the only manifestation of the aliphatic structures
of the UIE emitters.
%
%
They hypothesized that the clustering
of aromatic rings may break up the simple methyl- or
methylene-like side groups and hence the aliphatic
components may take many other forms.
They speculated that the in-plane and out-of-plane
bending modes of these side groups may combine to
emit the broad ``plateau'' emission
around 8 and 12$\mum$.
%
We note that the PAH model naturally accounts for
the so-called ``plateau'' emission
through the combined wings of
the C--C and C--H bands.
Wagner et al.\ (2000) have experimentally shown
that the plateau emission underlying the 3.3$\mum$ feature
is a general spectral feature of vibrationally
excited PAHs containing aliphatic hydrogens,
especially those containing methyl groups.
The plateau emission may merely be
an artificial result of
decomposing the 3--20$\mum$
UIE spectra into three components:
the UIE bands, broad plateaus,
and a thermal continuum.
We also note that the clustering of aromatic rings
and aliphatic chains would be accompanied by
forming new C--C bonds and losing H atoms.
Laboratory measurements of coals
(similar to MAONs in structure)
have shown that
lowering
the H content leads to
aromatization (see Papoular 2001).

Finally, we note that in comparison with
the median ratio of $\langle \Iratio\rangle\approx 0.12$,
some protoplanetary nebulae have a much larger
$\Iratio$ ratio (see Figure~\ref{fig:Iratio}),
with the 3.4$\mum$ feature exceeding the 3.3$\mum$ feature
(i.e., ${\rm I}_{3.4}/{\rm I}_{3.3}$\,$\gtsim$\,1; Hrivnak et al.\ 2007).
These are atypical UIE sources:
their UIE spectra have most of the power emitted from two broad bands
peaking at $\simali$8$\mum$ and $\simali$11.8$\mum$,
while typical UIE spectra have distinctive peaks at 7.7, 8.6,
and 11.3$\mum$ (see Tokunaga 1997, Peeters et al.\ 2004).
This is probably related to the fact these protoplanetary
nebulae do not emit much UV starlight.
Indeed, the 3.4$\mum$ feature (relative to the 3.3$\mum$ feature)
appears stronger in regions illuminated by stars
with a lower effective temperature $\Teff$
(X.J. Yang et al.\ 2013, in preparation).
It is likely that in the more benign, UV-poor environment
of protoplanetary nebulae, fragile species such as
aliphatic hydrocarbon chains produced in the outflow
could attach to an aromatic skeleton.
When evolving from the protoplanetary nebula phase
to the more hostile, UV-rich planetary nebula phase,
the aliphatic side chains are knocked off from the aromatic rings
as the aliphatic bonds are not as stable against
photodissociation as are aromatic bonds of normal PAHs.
As a result, PAHs with aliphatic chains would not be
expected to be numerous in UV-rich regions such as
planetary nebulae and the ISM. In contrast, they could
be present, perhaps even abundant, in more benign environments
(e.g., protoplanetary nebulae).
%

\section{Conclusion}\label{sec:summary}
We have examined the nature of the UIE emitters
based on the 3.3$\mum$ aromatic C--H emission feature
and its associated satellite features
at 3.4$\mum$ and beyond, with special attention
paid to the structure of the UIE carriers:
are they mainly aromatic or largely aliphatic
with a mixed aromatic/aliphatic structure?
The major results are:
\begin{enumerate}
\item We have compiled and analyzed the UIE spectra
      of 35 sources available in the literature
      which exhibit both the 3.3$\mum$
      and 3.4$\mum$ C--H features.
      We have derived $\Iratio$, the ratio of the power
      emitted from the 3.4$\mum$ feature to that from
      the 3.3$\mum$ feature for all these sources.
      With a median ratio of
      $\langle\Iratio\rangle\approx 0.12$,
      the majority (31/35) of these sources
      has $\Iratio < 0.25$.
\item We have computed $\Aali/\Aaro$,
      the band-strength ratio of
      the 3.4$\mum$ aliphatic C--H feature
      to that of the 3.3$\mum$ aromatic C--H feature,
      for a range of methyl-substituted PAHs,
      based on density functional theory.
      For these molecules,
      the $\Aali/\Aaro$ ratios all
      exceed $\simali$1.4,
      with an average ratio of $\Aali/\Aaro\approx 1.76$.
\item By attributing the 3.4$\mum$ feature
      {\it exclusively} to aliphatic C--H stretch
      (i.e., neglecting anharmonicity and superhydrogenation),
      we derive the fraction of C atoms
      in aliphatic form from $\Iratio\approx 0.12$
      and $\Aali/\Aaro\approx 1.76$ to be $\sim$\,2\%.
      We conclude that the UIE emitters are
      predominantly aromatic.
\end{enumerate}

\acknowledgments{%
We thank the anonymous referee for helpful suggestions.
AL and XJY are supported in part by
NSF AST-1109039, NNX13AE63G, NSFC\,11173019,
NSFC\,11273022, and the University of Missouri Research Board.
RG is supported in part by NSF-PRISM grant
Mathematics and Life Sciences (0928053).
Computations were performed using the high-performance computer
resources of the University of Missouri Bioinformatics Consortium.
}


\end{document}